\documentclass[twocolumn,showpacs,superscriptaddress,amsmath,amssymb,prb]{revtex4-1}

\usepackage{graphicx}
\usepackage{dcolumn}
\usepackage{bm}

\begin{document}

\title{Pressure dependence of the magnetic order in CrAs: a neutron diffraction investigation}

\author{L.\,Keller}\email{lukas.keller@psi.ch}
\author{J.\,S.\,White}
\author{M.\,Frontzek}
\affiliation{Laboratory for Neutron Scattering and Imaging, Paul Scherrer Institut, CH-5232 Villigen PSI, Switzerland}

\author{P.\,Babkevich}
\affiliation{Laboratory for Quantum Magnetism, \'{E}cole Polyt\'{e}chnique F\'{e}d\'{e}rale de Lausanne, CH-1015 Lausanne, Switzerland}

\author{M.\,A.\,Susner}
\author{Z.\,C.\,Sims}
\author{A.\,S.\,Sefat}
\affiliation{Materials Science and Technology Division, Oak Ridge National Laboratory, Oak Ridge, TN 37831-6114, USA}
 
 \author{H.\,M.\,R\o nnow}
\affiliation{Laboratory for Quantum Magnetism, \'{E}cole Polyt\'{e}chnique F\'{e}d\'{e}rale de Lausanne, CH-1015 Lausanne, Switzerland}

\author{Ch.\,R\"uegg}
\affiliation{Laboratory for Neutron Scattering and Imaging, Paul Scherrer Institut, CH-5232 Villigen PSI, Switzerland}
\affiliation{Department of Condensed Matter Physics, University of Geneva, CH-1211 Geneva, Switzerland}
\date{\today}
\begin{abstract}
The suppression of magnetic order with pressure concomitant with the appearance of pressure-induced superconductivity was recently discovered in CrAs. Here we present a neutron diffraction study of the pressure evolution of the helimagnetic ground-state towards and in the vicinity of the superconducting phase. Neutron diffraction on polycrystalline CrAs was employed from zero pressure to 0.65 GPa and at various temperatures. The helimagnetic long-range order is sustained under pressure and the magnetic propagation vector does not show any considerable change. The average ordered magnetic moment is reduced from 1.73(2) $\mu_B$ at ambient pressure to 0.4(1) $\mu_B$ close to the critical pressure \emph{P$_c$}$\approx$0.7 GPa, at which magnetic order is completely suppressed. The width of the magnetic Bragg peaks strongly depends on temperature and pressure, showing a maximum in the region of the onset of superconductivity. We interpret this as associated with competing ground-states in the vicinity of the superconducting phase. 

\end{abstract}


\maketitle
The magnetic structure of chromium mono-arsenide was first investigated more than four decades ago.\cite{Watanabe69, Selte71, Boller71} At \emph{T$_N$}$\approx$265 K and ambient pressure CrAs undergoes a first-order phase transition to a helimagnetic state. The magnetic propagation vector \textbf{k} is found to be parallel to the $c$-axis and the magnetic moments lie in the $ab$ plane. The magnetic structure can be seen as a set of four magnetic spirals along $c$, one for each Cr in the crystallographic unit cell, with well-defined magnetic phase angles between the magnetic spirals. This first-order magnetic phase transition is accompanied by abrupt changes of the lattice parameters, most prominently seen by a sudden expansion of $b$ below \emph{T$_N$}.\cite{Boller71} Nevertheless the orthorhombic MnP-type crystal structure with the space group \emph{Pnma} is preserved in the full temperature range.

The discontinuous onset of magnetic order has been related to a transition between collective and localized electronic states, with the abrupt change of $b$ as an important feature of this transition.\cite{Boller71} Such a system with coupled structural, magnetic and electronic properties is expected to be sensitive to pressure or doping which motivated the investigation of the CrAs system under pressure. First pressure studies on CrAs reported that \emph{T$_N$} shifts to lower values with pressure and vanishes completely at ~0.45 GPa.\cite{Zavadskii80} Very recently, the pressure-temperature phase diagram of CrAs has been independently investigated by Wu et al.  \cite{Wu14} and Kotegawa et al.  \cite{Kotegawa14a} using resistivity measurements under pressure. These studies each confirm that the magnetic ordering temperature \emph{T$_N$} drastically decreases with pressure and that the magnetic order is completely suppressed above a critical pressure \emph{P$_c$}$\approx$0.7 GPa. Remarkably, superconductivity was discovered to appear on suppression of the magnetic phase, displaying a maximum superconducting transition temperature \emph{T$_c$}$\approx$2.2 K at about 1 GPa.\cite{Wu14, Kotegawa14a} Increasing the pressure further decreases \emph{T$_c$}, and the superconducting phase adopts a dome-like shape. This pressure-temperature phase diagram with the gradual suppression of the magnetic state and the superconducting dome resembles that of many superconducting systems where unconventional superconductivity is induced by pressure or chemical doping.\cite{Jin11, Paglione10}

Wu et al. reported the onset of superconductivity already at $\sim$0.3 GPa and a gradual increase of the superconducting volume fraction up to \emph{P$_c$}, i.e. there is a region of coexistence of the magnetic and superconducting states. A very recent nuclear quadrupole resonance study under pressure\cite{Kotegawa14b} on CrAs reported that the internal field in the helimagnetic state only decreases slowly with increasing pressure, but maintains a large value close to \emph{P$_c$}. This indicates that the pressure-induced suppression of the magnetic order is of first-order. Therefore, even though substantial fluctuations are present in the paramagnetic state, the system is not close to quantum criticality.\cite{Kotegawa14b}

An incommensurate structure as found in CrAs at ambient pressure may be the result of competing exchange interactions or nesting. Such a structure is expected to be sensitive to distortions introduced by additional electronic or magnetic processes. A region of coexistence of magnetism and superconductivity is also a region of competing ground-states and the magnetic structure might be distorted. Features of the magnetic structure close to or in the region of coexistence may therefore act as a sensitive local probe. Neutron scattering is a direct and bulk microscopic probe of magnetic order and domains, which has not yet been studied in CrAs in the vicinity of the pressure-induced superconducting phase. Here we report neutron powder diffraction experiments on the magnetic ground-state of CrAs towards and in the vicinity of the superconducting phase.

\begin{figure}
\includegraphics[width=8cm]{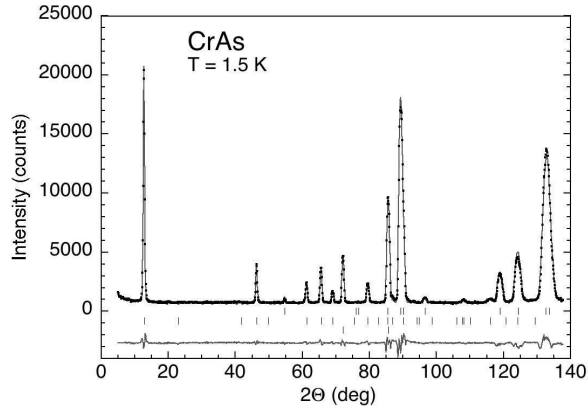}
\caption{\label{fig1}Refinement of the neutron powder diffraction pattern of CrAs at \emph{T}=1.5 K and ambient pressure. The three phases used for the refinement are the crystal structure of CrAs, the magnetic structure of CrAs and crystal structure of NaCl, neutron wavelength 3.804 \AA.}
\end{figure}

The polycrystalline sample of CrAs was synthesized as previously described.\cite{Saparov12} The neutron powder diffraction measurements were carried out using the cold neutron powder diffractometer DMC at the Swiss Spallation Neutron Source SINQ, Paul Scherrer Institute, Switzerland. The wavelength of the neutron beam was 3.804 \AA. For the reference measurement at ambient pressure, CrAs powder was enclosed in a V container and measured at 1.5 K, 80 K and 300 K. The clamp pressure cell was mounted in a He cryostat, covering a pressure range from ambient pressure up to 0.65 GPa at temperatures between 1.2 K and 300 K. Within the pressure cell the powder sample was enclosed in a lead capsule and a fluorocarbon-based fluid was used as pressure-transmitting medium. The applied pressure was determined by measuring the change of lattice constant of NaCl mixed with the sample. Profile refinements of the powder diffraction patterns were performed using the Rietveld software package FullProf Suite.\cite{Rodriguez93}

As a reference for the experiments under pressure we performed neutron powder diffraction measurements on CrAs (mixed with NaCl as pressure calibrant) at ambient pressure in the paramagnetic state at \emph{T}=300 K and in the magnetically ordered state at \emph{T}= 80 K and 1.5 K. The ambient pressure diffraction pattern of CrAs at 1.5 K is presented in Fig. 1. Also shown are the calculated patterns from the refined structural and magnetic models and the difference of observed and calculated pattern, showing the excellent agreement of experimental data and model calculation. The three refined phases are the crystal and helimagnetic structures of CrAs and the crystal structure of NaCl. The analysis of the data confirmed the type of magnetic structure found in literature \cite{Selte71, Boller71}, i.e. incommensurate helical magnetic order. The magnetic propagation vector \textbf{k} is of the form \textbf{k}=(0,0,\emph{k$_c$}) and the ordered moments lie in the ab plane (Fig. 2). The refined ordered magnetic moment per Cr is 1.73(2) $\mu_B$, the non-zero component of the magnetic propagation vector \emph{k$_c$}= 0.3562(2) and the magnetic phase angle $\phi$=-110(4)¡, using the notation introduced in Ref. 2. $\phi$ deviates from previously published values, which cannot reproduce our data. In the earlier studies the strongest magnetic peak (00\emph{k$_c$}) was dominated by a large background at low angles which flawed the determination of its integrated intensity. In our experiment we avoided this problem by an experimental set-up with long neutron wavelength of 3.804 \AA.

\begin{figure}
\includegraphics[width=4.5cm]{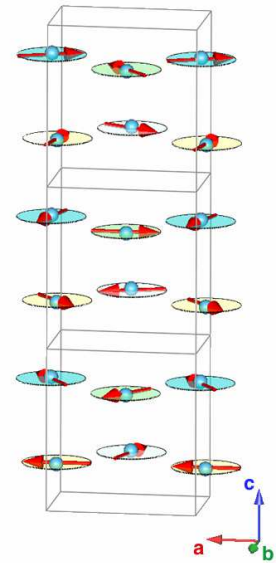}
\caption{\label{fig2}(color online) Incommensurate helical magnetic structure of CrAs. The evolution of the moments is shown for three unit cells along $c$; the four spirals are marked in individual colors.}
\end{figure}

\begin{figure}
\includegraphics[width=8cm]{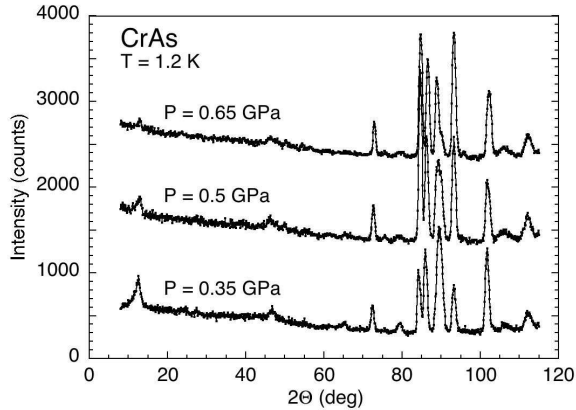}
\caption{\label{fig3}Pressure dependence of the neutron diffraction patterns at T = 1.2 K for the pressures \emph{P}=0.35, 0.5 and 0.65 GPa. The intensity axis relates to the 0.35 GPa data; other data sets have been shifted for clarity.}
\end{figure}

\begin{figure}
\includegraphics[width=6cm]{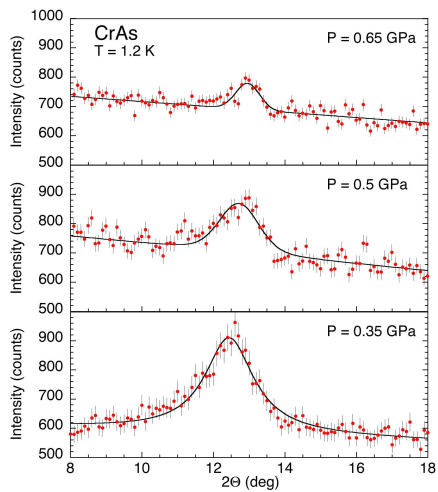}
\caption{\label{fig4}(color online) Pressure dependence of the first magnetic satellite (00$\pm$\emph{k$_c$}) for the pressures \emph{P}=0.35, 0.5 and 0.65 GPa. A convolution of Lorentzian and Gaussian peak shape functions was used for the profile fits.}
\end{figure}

\begin{figure}
\includegraphics[width=8cm]{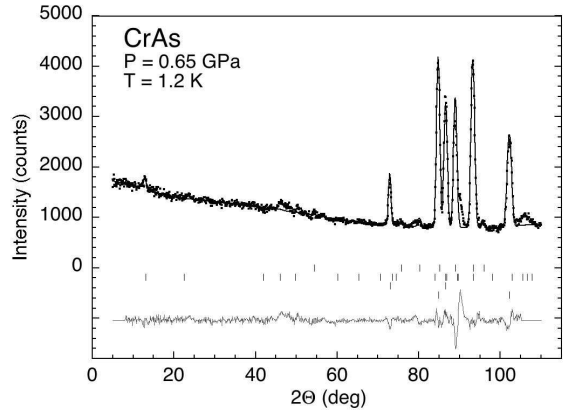}
\caption{\label{fig5}Refinement of the neutron powder diffraction pattern of CrAs at \emph{T}=1.2 K and P = 0.65 GPa, including the crystallographic phase of CrAs (Le Bail fit), the helimagnetic phase of CrAs, NaCl as pressure calibrant and Pb used for sample capsule in the pressure cell, neutron wavelength is 3.804 \AA.}
\end{figure}

The pressure dependence of the neutron diffraction patterns at base temperature of 1.2 K, measured at 0.35 GPa, 0.5 GPa and 0.65 GPa is shown in Fig. 3. Analysis of the data revealed that no crystallographic or magnetic phase transition occurred and that the symmetries of the crystal and magnetic lattices are preserved. Besides the peak shifts due to shrinking of the lattice with pressure there are clear changes in intensities of crystallographic Bragg peaks. While we are able to partially account for these intensity changes by refining atomic coordinates, the limited number of measured peaks does not allow for a full crystallographic refinement. Therefore the crystallographic phase was treated using the profile matching Le Bail method.\cite{LeBail88} The magnetic intensities decrease with pressure, as also seen in Fig. 4, which could indicate a reduction of the average ordered moment. No clear evidence was found for a reorientation of the magnetic moments, which is in agreement with recent NMR and NQR measurements.\cite{Kotegawa14b} Therefore for the refinement of the magnetic structure the zero pressure helimagnetic model was assumed, with the ordered moment and the propagation vector as refinable parameters. As an example Fig. 5 shows the profile refinement of the diffraction data at \emph{T}=1.2 K and \emph{P}=0.65 GPa, including the crystallographic phase of CrAs (Le Bail fit), the helimagnetic phase of CrAs, NaCl as pressure calibrant and Pb used for the sample capsule in the pressure cell. The magnetic peaks of CrAs show a clear change in width and shape (Fig. 4 for (0,0,$\pm$\emph{k$_c$})), which is not seen for either the crystallographic CrAs peaks or the NaCl peaks. Therefore, an inhomogeneity induced by the pressure transmitting medium at low temperature can be ruled out. Peak widths and peak shapes are sensitive probes for distortions to the magnetic structure; this will be discussed below.

The results of the magnetic refinements are summarized in Figs. 6(a) and 6(b). Fig. 6(a) shows the pressure dependence of the average ordered Cr moment for the helimagnetic structure model. The average moment decreases with pressure to 0.38(10) $\mu_B$ at 0.65 GPa, close to the critical pressure \emph{P$_c$}=0.7 GPa for total suppression of magnetic order. The pressure dependence of the internal field \emph{H$_{int}$} in CrAs shows a first-order transition to the paramagnetic state at \emph{P$_c$}.  \cite{Kotegawa14b} This behavior as well as the suppression of magnetic fluctuations at low temperatures \cite{Kotegawa14b} suggest that the suppression of magnetic order as function of pressure does not end in a magnetic quantum critical point at P$_c$. The moment deduced from \emph{H$_{int}$} (Fig. 6(a)) shows only a moderate decrease with pressure as compared to the neutron results. Neutron diffraction measures the whole sample while NMR probes the magnetic domains only. Therefore, the ordered moment deduced from neutron diffraction is averaged over the whole sample. The observed discrepancy can be explained with a reduced magnetic volume fraction and the refined size of the ordered moments with the sample under pressure should be regarded as a lower bound. Together with the observation by Wu et al. of a gradual increase of the superconducting volume fraction with increasing pressure\cite{Wu14} our diffraction results point towards a large coexistence region of magnetic order and superconductivity in CrAs.

Independent of the magnetic volume fraction, the magnetic propagation vector shows only marginal pressure dependence within the error of the fit (Fig. 6(b)), i.e. the periodicity or turning angle of the magnetic spirals remains stable with applied pressure. This is a remarkable behavior since the incommensurate magnetic structure in CrAs is regarded to be the result of competing exchange interactions \cite{Boller71} and one would expect some response to the changes in atomic distances and angles caused by applied pressure. On the other hand \textbf{k} is also very stable as a function of temperature at ambient pressure and the present magnetic order seems clearly favored. 

\begin{figure}
\includegraphics[width=7cm]{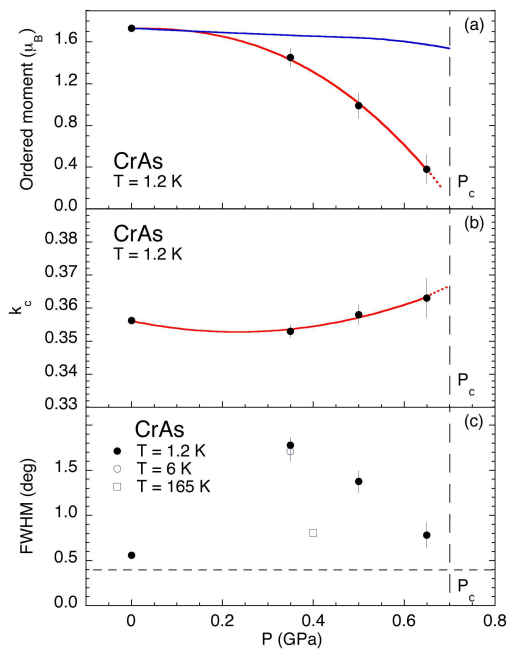}
\caption{\label{fig6}(color online) Pressure dependence of magnetic and peak shape parameters for \emph{P}=0, 0.35, 0.5 and 0.65 GPa at \emph{T}=1.2 K. (a) average magnetic moment for the helimagnetic model (red) and deduced from \emph{H$_{int}$} in Ref. 9 (blue), (b) non-zero component of the magnetic propagation vector, (c) full width at half maximum FWHM for the magnetic satellite (00$\pm$\emph{k$_c$}); the horizontal dashed line denotes the instrumental peak width.}
\end{figure}

The pressure dependence presented in Fig. 4 showed not only a reduction of intensity due to a reduced magnetic moment, but also a change in peak shape. Therefore it is interesting to also investigate temperature dependent effects at a fixed pressure. The temperature dependence of the diffraction patterns at 0.35 GPa is shown in Fig. 7. Again, we are able to partially account for changes of crystallographic peak intensities by refining atomic coordinates, but a full crystallographic analysis within the pressure-temperature phase diagram will be subject to future investigations. The temperature evolution of the magnetic (00$\pm$\emph{k$_c$}) peak is shown in Fig. 8. The refinement reveals that the ordered magnetic moments at 1.2 K and 6 K are the same within the error. But as for the pressure dependence at 1.2 K (Fig. 4) the width and the shape of the magnetic (00$\pm$\emph{k$_c$}) peak shows drastic changes. The peak shape function used for the profile fits in Fig. 8 is a convolution of a Lorentzian and Gaussian function with the respective weights as a free parameter. For the magnetic peak shape at 1.2 K we obtain a pure Lorentzian function, and at 6 K a mixed Lorentzian/Gaussian with equal weight and at 165 K an almost pure Gaussian function, which reflects the instrumental peak shape. For constant-wavelength diffraction used in the experiments presented here, Lorentzian peak shape is a textbook example of broadening due to size effects or due to reduction or distribution of domain sizes. 

The pressure and temperature evolution of the full width at half maximum (FWHM) of the magnetic (00$\pm$\emph{k$_c$}) peak is plotted in Fig. 6(c). The FWHM shows a maximum at 1.2 K and 0.35 GPa, just in the region of coexistence of magnetism and superconductivity.\cite{Wu14} Increasing the pressure leads to a reduction of the FWHM and at 0.65 GPa it is close to the zero pressure value. The same observation applies for increasing temperature at 0.35 GPa. The peak width is determined by instrumental parameters as well as by physical properties of the sample. The instrumental set-up was not changed during the measurements and also an artifact due to freezing of the pressure transmitting medium can be excluded as this would not only affect the magnetic peaks. Therefore the observed magnetic peak broadening is an intrinsic property. The peak width of magnetic peaks reflects the magnetic correlation length or domain size. While the exact origin of the peculiar behavior of the FWHM cannot be determined from our experiments, it is worth noting that the maximum width is seen in the region of the onset of superconductivity as determined by Wu et al. \cite{Wu14}

\begin{figure}
\includegraphics[width=8cm]{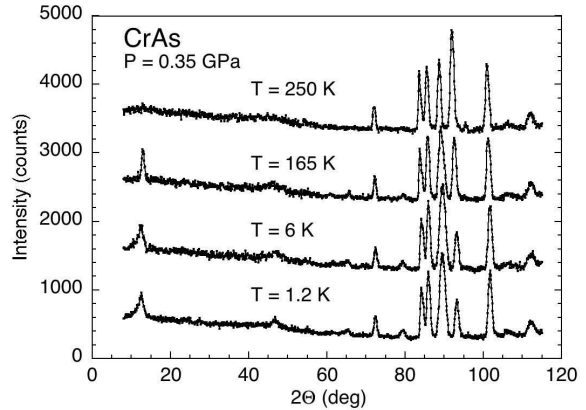}
\caption{\label{fig7}Temperature dependence of the neutron diffraction patterns for \emph{T}=1.2 K (0.35 GPa), 6 K (0.35 GPa), 165 K (0.40 GPa) and 250 K (0.43 GPa). The intensity axis relates to the 1.2 K data, the other data sets have been shifted for clarity.}
\end{figure}

\begin{figure}
\includegraphics[width=6cm]{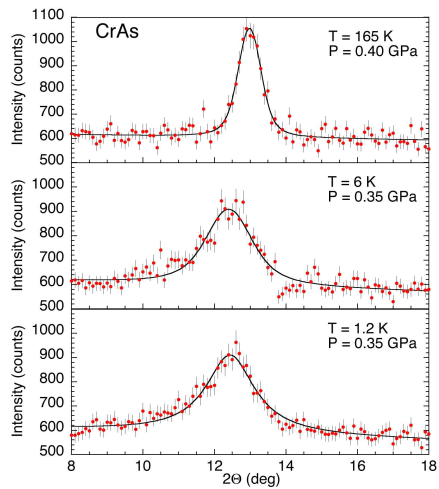}
\caption{\label{fig8}(color online) Temperature dependence of the first magnetic satellite (00$\pm$\emph{k$_c$}) for the temperatures \emph{T}=1.2, 6, 165 and 250 K. A convolution of Lorentzian and Gaussian peak shape functions was used for the profile fits.}
\end{figure}

The transition to the magnetic state in CrAs is interesting since it is first-order and changes the system from a Pauli paramagnet to an incommensurate antiferromagnet \cite{Zavadskii80}  with an equal moment spiral. Additionally, the incommensurate wavevector does not display a significant pressure- or temperature-dependence. This indicates that the magnetic order in CrAs might be due to Fermi surface nesting, with the magnetic propagation vector determined by a nesting vector. When the Fermi surface or the density of states at the Fermi energy is significantly changed, either through temperature or pressure, so that the nesting condition is no longer fulfilled, the magnetic order is spontaneously destroyed.  While we do not want to speculate on the pairing mechanism in CrAs, the properties in the region of co-existing antiferromagnetism and superconductivity might indicate a common origin for both ordering phenomena. The expected change of the Fermi surface under pressure ultimately promotes paramagnetic fluctuations which in a weak-coupling theory then can lead to an unconventional pairing formalism.\cite{Monthoux92} The onset of the superconducting phase marks the beginning of competition in CrAs where the electronic system either fulfills the Fermi nesting condition or the Cooper pair formation through paramagnetic fluctuations. At this point one can expect an energy equilibrium between both states, leading to pronounced fluctuations between these two possibilities. In our experiment, this manifests as strong broadening of the magnetic peaks. On driving the system further into the superconducting state by increasing the pressure, the superconducting state becomes the ground state and the fluctuations between magnetic and superconducting state become gapped. Consequently, we speculate that this gapping of the magnetic fluctuations leads to the counterintuitive reduction of the magnetic peak half-widths observed in our experiments on the approach to \emph{P$_c$}. The same argument can be used to explain the decrease of the magnetic half-widths with increasing temperature. However, in this case, the electronic system has no second state to occupy and therefore fluctuations are suppressed.

In summary, we have investigated the pressure and temperature evolution of the helimagnetic order in CrAs. The magnetic periodicity shows only marginal changes with pressure up to 0.65 GPa. The ordered magnetic moment averaged over the whole sample is reduced from 1.73(2) $\mu_B$ at ambient pressure to 0.38(10) $\mu_B$ close to the critical pressure \emph{P$_c$}, indicating a large coexistence region of magnetic order and superconductivity. An unusual behavior of the magnetic peak width and peak shape has been measured. We propose that its distinct pressure dependence can be interpreted as a fingerprint of underlying fluctuations  of magnetic and superconducting ground-states.

Part of this work was performed at the Swiss Spallation Neutron Source SINQ, Paul Scherrer Institute, Villigen, Switzerland. The work in Switzerland was supported by the Swiss National Science Foundation, its Sinergia network MPBH and the European Research Council project CONQUEST. The work at Oak Ridge National Laboratory was supported by the Department of Energy, Basic Energy Sciences, Materials Science and Engineering Division; also by the LDRD program.


\bibliography{Keller_CrAs}

\end{document}